# (Computer) Vision in Action: Comparing Remote Sighted Assistance and a Multimodal Voice Agent in Inspection Sequences

Damien Rudaz

Department of Nordic Studies and Linguistics, University of Copenhagen, Copenhagen, Denmark, daru@hum.ku.dk

Barbara Nino Carreras

Department of Nordic Studies and Linguistics, University of Copenhagen, Copenhagen, Denmark, bnc@hum.ku.dk

Sara Merlino

Department of Nordic Studies and Linguistics, University of Copenhagen, Copenhagen, Denmark, sme@hum.ku.dk

Brian L. Due

Department of Nordic Studies and Linguistics, University of Copenhagen, Copenhagen, Denmark, bdue@hum.ku.dk

Barry Brown

Department of Computer Science, University of Copenhagen, Copenhagen, Denmark, barry@di.ku.dk

Does human–AI assistance unfold in the same way as human–human assistance? This research explores what can be learned from the expertise of blind individuals and sighted volunteers to inform the design of multimodal voice agents and address the enduring challenge of proactivity. Drawing on granular analysis of two representative fragments from a larger corpus, we contrast the practices co-produced by an experienced human remote sighted assistant and a blind participant—as they collaborate to find a stain on a blanket over the phone—with those achieved when the same participant worked with a multimodal voice agent on the same task, a few moments earlier. This comparison enables us to specify precisely which fundamental proactive practices the agent did not enact in situ. We conclude that, so long as multimodal voice agents cannot produce environmentally occasioned vision-based actions, they will lack a key resource relied upon by human remote sighted assistants.

## 1 INTRODUCTION

Over the past decade, a lively debate has centered on whether and how algorithmic systems collaborate with humans [19,22,60,78,124,164], including the possibility of joint action in human-robot interaction (HRI) [18,163] or, more broadly, human-agent interaction [53,90]. This ongoing discussion questions whether human-'AI' interaction features "truly interactive" [60] practices constitutive of many forms of human collaboration. Consequently, a central point of contention in this debate is strictly empirical, rather than ontological [53,160]: namely, whether state-of-the-art conversational agents currently accomplish the subtle collaborative practices found between humans, with a comparable degree of proactive behavior and finely tuned adjustment. We confront this issue by comparing human and AI-powered assistance during the 'same' task.

Put differently, this study addresses and derives design insights from a deceptively simple question: *if, to an external observer, two humans coordinating over a smartphone remain intuitively distinguishable from an interaction between a human and a multimodal voice agent on the same device, then what practices make these two situations recognizably*

*different?* Through this inquiry, we aim to clarify which collaborative capacities today's multimodal voice agents do (and do not) exhibit in situated interaction. Our analysis builds on a detailed comparison of how a blind participant searched for a stain on a blanket, depending on whether her assistant was a *human remote sighted assistant* (RSA) or a *multimodal voice agent*. The results of this empirical analysis underscore the continued relevance of the notion of proactivity [20,37] for explaining the current limitations of multimodal voice agents in collaborative tasks. We detail specific practices—initiating or modifying joint action—that voice agents currently do not produce, and highlight, by contrast, how these practices were crucial to the successful coordination between human participants. In particular, we find that so long as multimodal voice agents cannot *initiate* actions in response to the video data available to them, they will lack a key resource routinely relied upon by human sighted assistants. Framed another way, we use an episode of human–voice agent interaction as a backdrop against which to contrast the specificities of human collaborative practices and, notably, to foreground the mechanisms on which these practices fundamentally rely. This allows us to reiterate some often-overlooked *conditions of possibility* for efficient artificial agents in embodied assistive tasks.

In doing so, we align with a growing body of literature that seeks to document the concrete interactional phenomena underlying technical terms (in this case, "remote sighted assistance" [97,192,193,197], "human-AI collaboration" [15,53,60,164], and "proactivity" [20,32,37,198]) in order to advance the design of voice agents [55,142,143] and access technologies [1,149,151,152,194]. Nevertheless, beyond technical and practical considerations, we conclude by questioning whether it is ethically appropriate for voice agents to reproduce the full range of proactive behaviors that characterize human collaboration. Indeed, our comparison demonstrates how, in joint activity, initiating new actions enforces a specific system of values about what is relevant to do and to attend to in a given situation.

## 2 BACKGROUND

### 2.1 Human and AI-Powered Assistance

#### 2.1.1 Remote Sighted Assistance

The following comparison is conducted regarding the domestic task of "spot-checking" [25]—namely, locating a stain on a blanket—in which a blind person cooperates with a sighted participant via a smartphone. Identifying the presence of spots or stains is a recurring activity in which blind and low vision (BLV) individuals frequently enlist the collaboration of a sighted partner; for example, to provide final confirmation about the absence of dirt on floors or stains on fabrics [25,153,181], particularly when the spots in question do not afford tactile cues. Cleaning and organizing tasks are well-documented challenges for BLV people [148], and are often described as "a common source of frustration" [181]. Crucially, in the absence of physically co-present partners, BLV individuals can now rely on applications offering remote sighted assistance services [97,195]. These applications connect BLV users to sighted volunteers who gain access to their immediate surroundings through the BLV person's smartphone camera and microphone. As a pair, both participants can then collaborate to co-accomplish tasks such as verifying the presence of dirt or stains. In other words, BLV individuals are often engaged in "inspection sequences" [50,126] (stretches of interaction where participants closely examine their surroundings or a particular object) that involve a sighted participant, whether physically co-present or remote.

#### 2.1.2 AI-Powered Assistance

Increasingly, however, BLV people are also turning to AI-powered remote assistants rather than human volunteers [1,35,63,64,188]. Indeed, in recent years, the advancement of visual description applications enabled by multimodal large language models (which process photographs and videos as inputs) has offered new ways of accessing visual information [7,194]. Yet, because specialized assistive tools for cleanliness-related tasks remain relatively rare [153,181], these AI-powered assistants are now often commercially available multimodal voice agents such as ChatGPT multimodal "voice mode", Google Gemini Live, or Meta AI [4,31,179,194]. These agents rely on multimodal language models that can interact in natural language with BLV people while also generating responses based on visible features of the environment, as captured through a device's camera and microphone (e.g., a smartphone or smart glasses). They can be used by BLV people to detect and discretize observable features of their environment (such as dirt on the floor), or to receive descriptive overviews of their surroundings [30,31].

Consequently, these artificial assistants must be able to *co-accomplish* assistive tasks with BLV individuals—for example, locating a stain on a blanket. Yet, as noted by the few existing studies that have examined multimodal agents in real-world scenarios [31,194], these agents still frequently fail to accomplish these assistive tasks reliably. These difficulties extend across the entire field of applications that provide visual descriptions of the environment—whether specialized tools (e.g., VisionPal [64]) or voice agents. For instance, Gonzalez Penuela et al. [63] found that BLV people were frequently skeptical of and dissatisfied with visual description applications powered by multimodal LLMs. A year later, Gonzalez Penuela et al. [64] conducted a two-week diary study with 20 BLV people focusing on one specific LLM-powered visual description application. In contrast to earlier findings, they reported preliminary results showing that participants were generally satisfied with the descriptions provided. However, they also noted that, despite higher trust rates, the application's output was at times erroneous or misleading. Moreover, they report that as trust increased, users also began using this application in high-stakes situations, such as inspecting medical records. Accordingly, the authors encourage studies that attend not only to users' trust but also to matters of accuracy and error. Similarly, Alharbi et al. [7] documented how, despite the usefulness of LLM-powered applications as visual descriptors, issues persisted with fabricated or inaccurate descriptions. In addition, they highlighted that the absence of real-time feedback (on what the smartphone's camera was capturing) remained an unmet expectation for BLV users of these applications.

## 2.2 Leveraging Remote Sighted Assistance to Improve AI-Powered Assistance

In this state of affairs, untapped potential for improvement remains. To date, very few studies have examined remote sighted assistance or AI-powered assistance from a fine-grained perspective. We still lack detailed accounts of the concrete, minute practices covered by the umbrella terms "remote sighted assistance" [97], "Human-AI Collaboration" [60,98,160], or "AI-Powered Visual Assistance" [197]. This gap has practical consequences: little is known about the forms of joint action that these terms encompass and that could inform potential improvements for voice agents. Notably, although several reviews and theoretical discussions have addressed proactivity as a useful concept and direction [20,36,37], it remains unclear what proactivity consists of in terms of the concrete, local, collaborative practices that a voice agent should produce in a given task.

RSA thus provides particularly favorable conditions for evaluating how far voice agents still have to go to achieve expert-level collaboration with humans: in RSA, the human assistant is embodied in the local setting through the same mediating object (e.g., a smartphone or smart glasses) as the AI-powered assistant. By itself, this material artifact offers, a priori, comparable affordances and resources to BLV persons, whether the task is co-accomplished with an AI-powered assistant or a human. In particular, in RSA or AI-powered assistance, "the instructors cannot use tactile resources or other locally situated sensory input such as smelling/feeling/sensing when assisting the activity" [49]. Likewise, both human assistants and AI agents receive visual and auditory access to this local setting through the same hardware—namely, the same microphones and cameras.

## 2.3 Ethnomethodological Conversation Analysis and Ordinary Expertise

Building on this comparability, the specialized practices of blind participants and remote sighted assistants appear to be a valuable resource for improving the effectiveness of multimodal voice agents in certain assistance tasks. In particular, it is now widely recognized that blind people and their co-participants display a wealth of under-documented practical knowledge [11,46,47,128,150,186]. To investigate this expertise and how it takes shape in interaction, we adopt the granular perspective of multimodal ethnomethodological conversation analysis (EMCA). This analytic approach is typically leveraged to investigate the moment-by-moment production of participants' conduct in a given setting [66,161]. It relies on the detailed analysis of corpora of video-recorded naturalistic interactions [71,72,117] to understand how participants' actions (such as pointing to a feature of the environment [12], inspecting an object [47,50], or providing instructions [106,170]), are produced and adjusted in relation to the conduct of their co-participants [117,119]. These co-participants are usually humans but may also include artificial agents [52,115]. Because of its attention to detail, the analytic perspective of EMCA is especially well suited to studying the ordinary expertise [10,44] displayed by members of society: that is, the subtle "seen but unnoticed" [58] practices, methods, or strategies through which they accomplish specific activities [109].

Crucially, EMCA approaches are especially sensitive to the parameters by which participants *initiate* [176] and form [101] actions. This involves studying how conversationalists come to take a specific turn at a particular moment [159], notice features of their environment [87], display them in a particular way [105], or modify their utterance at a given point [65]. Put differently—and significantly for this study—EMCA's analytic tools make it possible to specify how participants produce forms of conduct often characterized as 'proactive'.

## 2.4 Multimodal Voice Agents and Proactivity

### 2.4.1 The Challenge of Proactive Agents

Over the past few years, a growing line of research in human-computer interaction (HCI) has emerged on the proactivity of conversational agents[1] [20,108] and robots [36]. This field of research tackles the enduring problem of establishing if, how, and under which conditions an artificial agent[2] should initiate new actions without being explicitly prompted by a human [91,132]. In this understanding, proactivity refers to the capacity of an artificial agent to autonomously and appropriately take initiative in interaction: for example, producing actions that anticipate user needs or making timely contributions that are not explicitly prompted by a human [20,36,187]. Broadly construed, proactivity therefore includes initiating interactions [20,134,147] but also encompasses any form of initiative taken, at any relevant moment, during an ongoing interaction [36,38,108,137]—as opposed to situations in which "agent acts only in response to a user input or prompt" [20]. As noted by Bérubé et al. [20], the proactivity framework partially overlaps with the notion of mixed-initiative interaction in HCI [41] and HRI research [84]. Both lines of research address the challenge of dynamically sharing initiative between a human and a system, where each can opportunistically take or relinquish control of the task or dialogue as needed in pursuit of a joint goal [9,74].

---

[1] In line with a portion of the HCI literature [32,89], we define voice agents (sometimes referred to as voice-based agents [59,155]) as a specific form of conversational agents. Conversational agents denote any interactive system capable of natural-language dialogue (text, voice, or multimodal), whereas a voice agent refers specifically to a conversational agent that operates through spoken interaction alone [189]. However, categorizing ChatGPT "voice mode" as a form of conversational agent—though simpler for aligning with the literature—has the disadvantage of presupposing that ChatGPT "voice mode" is conversational. Indeed, as other works have noted [51,79], the 'conversational' nature of interaction with an LLM-based voice agent should be treated as an analytic finding rather than a starting assumption.

[2] By artificial agent, we refer to any computer system—such as a robot, a virtual assistant, or software—capable of autonomous behavior [17] in its interaction with humans [56].

*2.4.2 Proactivity and Turn-Taking*

Given its analytical focus outlined above, the proactivity framework specifically connects with the practical problem of turn-taking [159,172]: that is, determining when a conversational agent should initiate a new turn [38,75,132] or contribute to a multiparty conversation [108], even if no human has spoken immediately beforehand. *Multimodal* agents able to respond to features of the observable environment have further complicated this design problem [132]. Any perceived feature of these agents' surroundings—within the vast amounts of audio and video data they process—can potentially occasion a new action from these agents: from humans' gestures and speech [96,199] to other salient environmental cues [54]. In short, as the number of parameters these agents can process has exploded, so too have the opportunities to initiate or modulate their actions in real time [146,184].

Research on proactivity and multimodality is closely tied to efforts to improve the efficiency of artificial agents in cooperating with humans in task-oriented activities [15,27,37,76,103]. Indeed, proactive behaviors—in particular, the ability to appropriately initiate or modify courses of action—are at the very foundation of human cooperation and joint action [140]. In this regard, studies on proactivity are directly linked to a substantial body of work in EMCA (see Section 2.3) that documents the ongoing mutual adjustments accomplished by humans in interaction [67,117,118]. Although these studies draw on a different theoretical vocabulary and framework, they describe in fine-grained detail the complexity of the multimodal resources mobilized by humans in collaborative activities to initiate or modify their actions [40,122,125]—and which are now partially perceptually available to multimodal voice agents.

## 3 STUDY APPROACH

### 3.1 Data Collection

*3.1.1 Overall Corpora*

The two following fragments are representative exemplars of two overarching corpora of video data. The first fragment is part of a corpus of 20 recorded interactions focused on the examination of objects, jointly produced by BLV and sighted participants (20 distinct BLV–sighted pairs). The second fragment comes from a corpus of 11 recorded episodes of unique BLV participants using access technologies such as smart glasses and smartphone-based RSA applications. However, irrespective of the generalizability of the phenomena documented in this article (see Section 7 for a discussion of the limits of our analytic approach), the following insights should be treated as local: they derive exclusively from a heuristic comparison of these two fragments. Additionally, although the fragments belong to different corpora (depending on whether the 'assistant' is human or artificial), they feature the same blind participant.

*3.1.2 Study Setup and Ethics*

Our analysis draws on a subset of ethnographic fieldwork from which these two fragments were selected: four hours of video material filmed by the first and second authors in collaboration with a blind participant pseudonymized as Laura. These recordings focused on Laura's use of various access technologies to accomplish routine tasks, both in her apartment and while navigating the street.

Laura is a Danish citizen who speaks fluent English. She is a regular user of large language models, and her professional work requires substantial familiarity with these technologies. Building on HCI scholarship that emphasizes centering and engaging disabled people in disability-related research [174], the researchers involved Laura in the research design process. In particular, they worked with her to determine which routine activities would be relevant to film and how best to capture them, in order to understand the affordances and constraints of various technologies in her everyday life. Therefore, researchers did not ask Laura to perform pre-defined tasks or follow a fixed structure when using a particular technology.

Throughout the study, the researchers provided in situ scene descriptions to indicate camera placement and notified Laura whenever recording began or ended. The data were collected with Laura's informed written consent, covering both the collection of personal data and the publication of images of herself recorded during this fieldwork. The data collection took place in Denmark and was approved by the researchers' university ethics committee. In line with the compensation rules of the researchers' university, Laura received a gift worth 500 Danish kroner.

### 3.2 Data Analysis and Transcription

*3.2.1 Data Sessions in EMCA Analysis*

The video material presented in this study was analyzed through a series of three data sessions. Data sessions are a recurring step in EMCA practitioners' analytic process [14,71]. They are organized around the detailed examination of selected interaction recordings. During these sessions, several researchers first review the transcripts of these recordings and correct any imprecision or error identified [14,71]. Subsequently, the researchers propose or contest [5] candidate characterizations of the actions accomplished by the recorded participants, as well as the features of the setting that participants orient to as relevant. In this sense, as noted in [139,169], conducting several data sessions with different researchers aims both to increase the scope and to enhance the reliability of the analysis. By "reviewing video recordings and transcripts of the phenomenon collaboratively multiple times" [88], researchers (1) strengthen their chances of identifying the practices, processes, or methods that are produced by, and relevant to, participants in situ [21]; and (2) reevaluate the adequacy of current analyses of these phenomena [5]. That is, data sessions are used to "test the reproducibility of previous discoveries" [21]—as a form of informal peer review [6]—and to discover "new phenomena" [21].

*3.2.2 Analytic Process*

The materials used in this article were examined through this conventional data session procedure:

1. *In an initial data session,* the first and second authors watched extended segments drawn from their recordings with Laura. They identified two episodes as potential instances of phenomena that were recurring in their broader corpus; these episodes are those analyzed in this paper (Fragment 1 and Fragment 2). As a follow-up to this first data session, the first and second authors transcribed these two episodes using multimodal transcription conventions (see Section 3.1.1 and Appendix). In addition, both authors produced image descriptions of still frames from the video in order to reflect further on the material and to ensure accessibility for BLV readers.
2. *A second data session* was then organized around the two selected episodes, with all five authors of this article participating. They refined the transcripts of the two fragments presented in this article, examining them alongside the corresponding video recordings. Each author was familiar with the broader corpora of recordings. The data session ended after the authors reached agreement on the witnessability of several interactional phenomena in these recordings—the phenomena detailed in this article.
3. *Finally, a third data session* was conducted, again involving all five authors. They reviewed the phenomena identified in the previous sessions once again and confirmed that all relevant spoken and embodied details of these interactions had been rigorously transcribed and synchronized. As is routinely done during data sessions, these episodes were also discussed in relation to already documented phenomena in the existing CA literature. This comparison aimed to determine whether the recorded activities constituted specific instances of broader interactional

phenomena—either already labeled in the literature or, at the very least, describable using EMCA's concepts and analytic tools.

### 3.2.3 Transcription Conventions

Participants' conduct was transcribed using Jeffersonian conventions for talk [83], and Mondadian multimodal transcription conventions for embodied conduct [117]. The details of these transcription conventions are provided in the Appendix. Jeffersonian transcription captures the fine-grained features of speech—such as pauses, intonation, overlap, and timing—essential for analyzing the sequential organization of interaction [165]. Mondadian conventions allow annotators to represent participants' gaze, gesture, posture, and other embodied actions in relation to talk, i.e., to synchronize those dimensions over the same temporal stream. Combined, Jeffersonian and Mondadian conventions therefore enable the analyst to account for how participants *coordinate* their spoken and/or embodied conduct, moment by moment [117,118].

In the following transcripts, the stream of talk is shown in bold type, while the embodied conduct is placed on the line(s) immediately below in regular characters. The stream of talk is taken as the point of reference [117] with which the embodied conduct is synchronized. This synchronization is achieved through the use of identical symbols—for example, "$"—positioned one above the other in the line transcribing the talk and in the corresponding line representing the embodied conduct. These symbols delimit the beginning or the end of embodied actions (e.g., a pointing gesture, a change in posture, a shift in gaze direction).

## 4 CO-INSPECTING A BLANKET WITH A SIGHTED VOLUNTEER

### 4.1 Overview of the Fragments

In the following fragments, Laura is studied while using ChatGPT multimodal "voice mode" [136] and the smartphone application *Be My Eyes* [16]. ChatGPT multimodal "voice mode" is an LLM-powered interface that allows users to interact through spoken dialogue while it processes visual and auditory inputs from a device's camera and microphone. *Be My Eyes* is a mobile application that connects blind or low-vision individuals with sighted volunteers via video call to provide real-time assistance with tasks facilitated by visual perception. Laura used these two applications consecutively during the same afternoon, in her apartment. This afternoon was devoted to a naturalistic data collection focused on Laura's skills and ordinary practices as a blind person (see Section 4.1.2). During this time, Laura was filmed in her everyday activities by two researchers, who were present in the same room as Laura throughout the following excerpts. One camera was positioned in the corner of the room, and the other was handheld by a researcher.

The two 'spot checking' episodes unfolded in the following order. The use of the ChatGPT multimodal "voice mode" [136] application was initiated by Laura, who wanted to test the capacities of its new multimodal feature on some tasks she regularly undertook. Shortly before the researchers arrived, she had spilled some liquid from her water bottle onto her bed and was unable to determine whether, and where, the liquid had stained her blanket. Laura then stated that she would take this opportunity to solve this ongoing problem (See Figure 1).

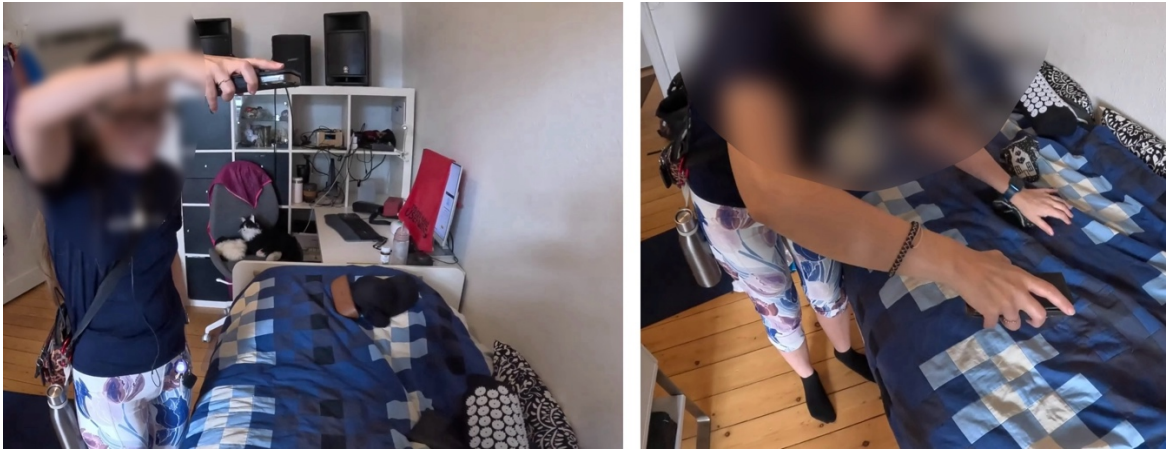

Figure 1. Laura unsuccessfully searches for a stain on her blanket using ChatGPT multimodal "voice mode"

When Laura's attempt to locate the stain using ChatGPT "voice mode" failed, Laura turned to the *Be My Eyes* application [16] to call a volunteer (See Figure 2). Laura did not know the identity of the volunteer who answered her call. However, she was already familiar with the *Be My Eyes* application, which she uses regularly. Significantly, this indicates that Laura has frequently collaborated with remote sighted assistants to accomplish various everyday tasks, and may have developed a form of expertise in doing so. Once the interaction with the *Be My Eyes* volunteer (pseudonymized as Morgan) was completed, Laura confirmed that they had successfully identified the location of the stain she was looking for.

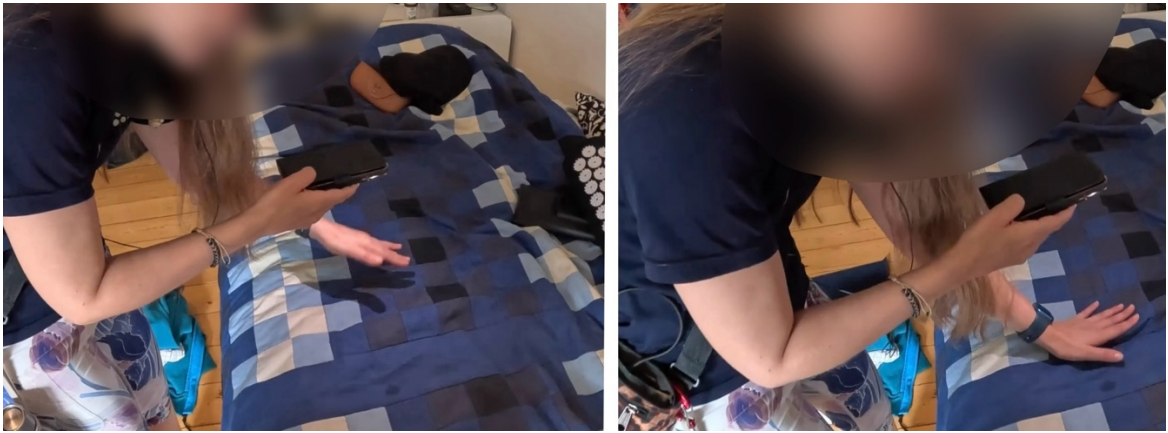

Figure 2. Laura finds the stain on the blanket with a *Be My Eyes* volunteer, Morgan.

The interaction between Laura and the *Be My Eyes* volunteer featured several practices that were entirely absent from the interaction between Laura and ChatGPT multimodal "voice mode". Not only did those practices make these fragments distinguishable, but they were also key to the eventual success of Laura and the volunteer in locating the stain. We examine these practices below.

## 4.2 Initiating and Modifying Turns in Response to Emergent Features of the Observable Setting

Across our data, a significant portion of the practices produced by the *Be My Eyes* volunteer (Morgan, abbreviated as MOR in the following transcripts) exhibit two properties that are entirely absent from the conduct of the voice agent:

1. **They are initiated in response to observable aspects of the setting.** The timing and design of several of Morgan's contributions signal that they are "occasioned" [87] by visual cues in the smartphone camera's field of view, rather than by Laura (LAU)'s talk.

An instance of turns occasioned by changes in the observable environment is evident in the following fragment (Fragment 1.1). This exchange occurs at the beginning of the joint inspection sequence. At this point, after Laura's request to help her find a stain on her blanket, Morgan has just stated that she can see the blanket but that she does not see any stain. Laura's left hand (that she uses to scan the bed) then enters the camera's field of view.

```
24      #((LAU's hand enters the camera's field of view))#
  fig #fig.3 (left)                                  #fig.3 (right)
25      (0.5)
26 MOR  feel arou:nd* (.) where do you feel the stain so i *can see- ok. (0.2) wait-
  lau            -->*-----stops scanning with her hand-----*scans again-->
27      (0.4)
28 LAU  e[h!]*
  lau   -->*rests hand on bed, out of the camera's field of view-->
29 MOR  [back]up a little bit?
```
Fragment 1.1. Morgan (MOR) constructs several turns that display responsiveness to Laura's (LAU) ongoing embodied course of action

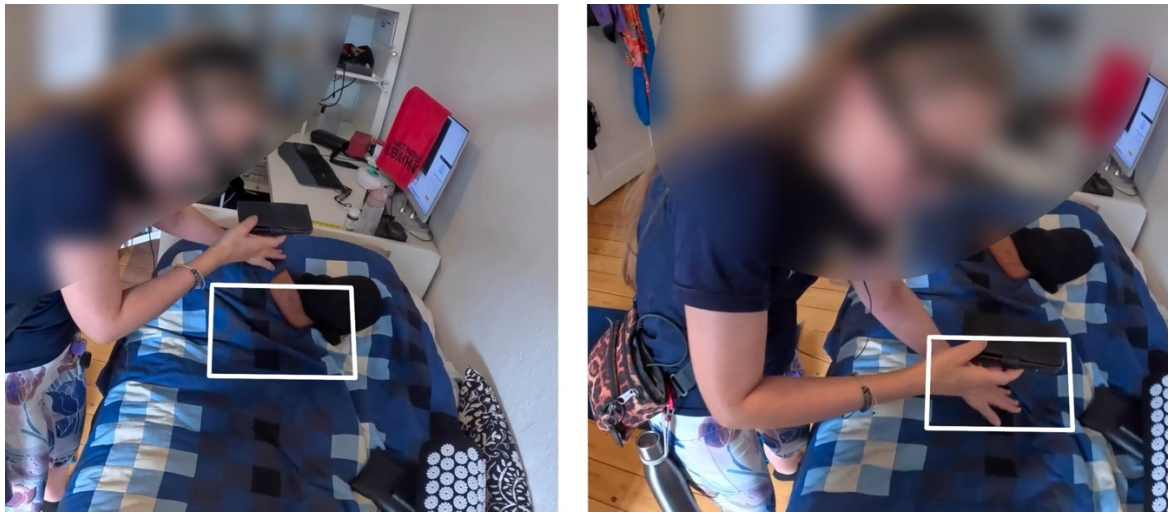

Figure 3. Laura's hand is scanning the bed outside of the camera frame (left), then enters the camera frame (right). Laura's hand appearing in the camera frame occasions an account from Morgan ("feel around", l.26).

Immediately after Laura's hand appears in front of the smartphone's camera, the volunteer utters "feel around" (l.26). This positions Morgan's contribution as an account of Laura's embodied action: it is constructed and timed as responsive to Laura's now visible hand movement. As such, it constitutes a resource available to Laura. It indicates what features of the setting Morgan currently perceives as relevant in the field of view of the smartphone's camera, and it displays her (positive) stance towards Laura's ongoing action.

2. **They are modified in real time,** in response to emergent features of the setting available through the smartphone's camera.

For instance, immediately after she utters "feel around" (l.26), Morgan produces two cut-off requests: first to "wait-" (l.26) and, then, to "backup a little bit" (l.29). This series of interruptions and restarts in Laura's utterance ("see-" then "wait-") displays its incremental construction in response to Laura's ongoing scanning movement with her hand on the bed. This turn from Morgan is only meaningful as a response to—and as being progressively modified by—the ongoing course of action from Laura available to Morgan through the smartphone's camera feed. Congruently, immediately after this turn, Laura removes her hand from the camera view (l.28).

Not only are Morgan (and Laura)'s actions occasioned by the perceived setting, they are also ongoingly and incrementally constructed *as they occur*. That is, in complete contrast with the conduct of ChatGPT "voice mode" in Fragment 2 (see Section 5), Laura and Morgan's interaction markedly relies on a mutual and online modification of their courses of action. Put differently, Laura and Morgan's actions are reflexively adjusted: each action builds on and modifies the other participant's ongoing action in real time, and vice versa [123].

### 4.3 Distributing the Perceptual Work

*4.3.1 Co-Producing a 'Good View'*

The preceding properties of Morgan's and Laura's contributions (occasioned and ongoingly adjusted) are central to their coordination around the bed. Morgan continuously instructs or displays to Laura which resources are relevant to her as a sighted volunteer. In short, Laura does not single-handedly initiate actions that modify the camera's perspective or that reconfigure the common workspace. When she accomplishes these moves, she does so by building on the resources provided by Morgan.

```
34      (0.8)*(01)
   lau   -->*scans bed in front of camera-->
35 LAU e::h it's a little wet (.) yeah. (0.2) heh.
36      (2.5)
37 MOR okay. (0.3)*#(0.8) let me see. (0.3) >backup a little bit- put your hand
   lau            -->*stops scanning, hand out of the camera's field of view-->
   fig              #fig.4 (left)
38      on the blanket<
39      (0.6)*#(0.4)
   lau   -->*puts hand flat on blanket, possibly out of the camera's field of view-->
   fig        #fig.4 (right)
```

Fragment 1.2. As Laura is scanning the blanket with her hand, Morgan gives several instructions in a row to Laura

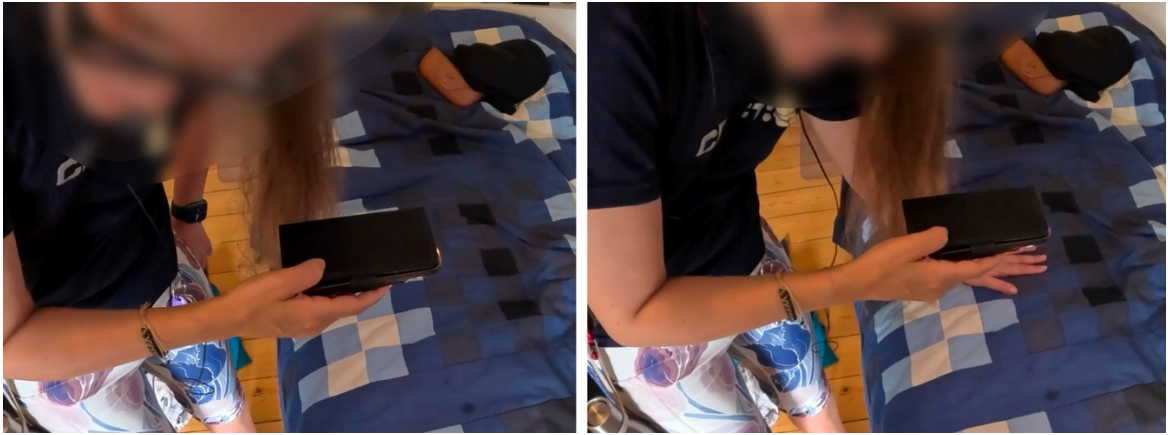

Figure 4. Laura points the smartphone camera towards the bed (left), then puts her hand flat on the blanket after Morgan instructs her to do so (right).

For instance, as Laura just stopped scanning the bed in front of the smartphone's camera (Fragment 1.2, l.37, Figure 4 left), Morgan requests Laura to "let her see" (l.37), to "backup a little bit" (l.37), and to provide her with a visual point of reference ("put you hand on the blanket", l.37 and l.38). Morgan therefore displays an orientation to reconfiguring the setting itself (through the addition of Laura's hand, l.39, Figure 4 right) rather than merely the perspective from which this setting is seen. Although Morgan does not directly accomplish these changes, she provides instructions aimed at securing a 'better' view (for the task at hand) that are, then, enforced through Laura's own actions. This conduct is strikingly different from that of ChatGPT multimodal "voice mode", which involved no such practices. Instead, in this fragment, Laura initiated all the rearrangements of the bed or camera's perspective, and treated none of ChatGPT's turns as useful resources in these rearrangements.

### 4.3.2 Co-Constructing Shared Reference Points

Morgan and Laura actively rely on and build upon each other's specific modal access to the setting (sighted or tactile). Their perceptions are rendered "complementary" as defined in [24,114]: each participant produces accounts of their own perceptual access to the setting that can be used as a resource by their interlocutor. This endeavor to calibrate and build upon each participant's specialized senses is encapsulated by Morgan's phrasing "where do you feel the stain so I can see" (Fragment 1.3, l.26), uttered as Laura's scanning motion becomes visible to Morgan. However, the practices through which Laura and Morgan aligned their senses were, for the most part, *instructed* by Morgan. That is, the complementarity of participants' sensorial experience was structured by several methods[3] initiated by Morgan to calibrate sight and touch.

One such method is displayed in Morgan's instructions to Laura in creating a common reference point (l.42 to l.48). The creation of a reference point in space is well documented in the literature on blind individuals: a reference point refers to any element of the environment (perceived through touch, haptics, or hearing) used to orient oneself in space [47,48,145]. In this case, however, the experience of the reference point is calibrated to create a correspondence between the sighted participant's vision and the blind participant's embodied experience of the environment.

---

[3] These methods may display Morgan's expertise as a remote sighted assistant on the *Be My Eyes* application; however, Morgan's exact background and level of familiarity with remote sighted assistance remain entirely unknown.

```
42 MOR  touch- touch the camera and then: <place: (.) your ha:nd> (.) right
43      beneath where the camera would be
44      (0.4)*#(0.4)
   lau  -->*puts hand below phone's camera-->
   fig       #fig.5 (left)
45 LAU  ok. (.)[so* (.)] there's [the (.) y]eah.
   lau         -->*-->puts hand flat on bed-->
46 MOR         [tsk. okay]      [perfect]
47 MOR  =okay. (.) now move your hand to the right
48      (0.4)*%#(0.6)
   lau  -->*moves hand right-->
   lau        %shifts body and camera to the right-->
   fig          #fig.5 (right)
```

Fragment 1.3. Creation of a common reference point, instructed by Morgan

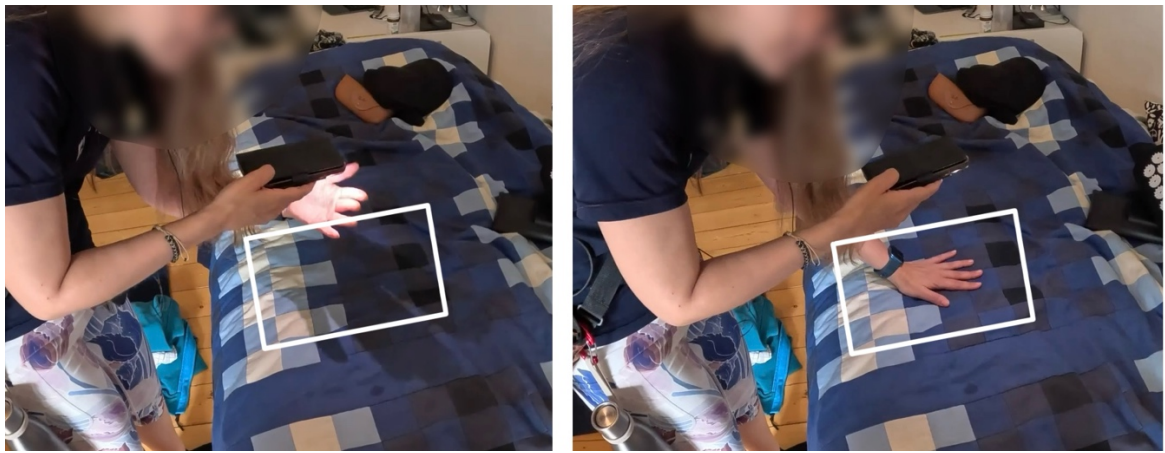

Figure 5. Laura puts her hand on the smartphone's camera (left), then on the bed directly below the camera (right). The rectangle represents the approximate field of view provided by the camera.

Morgan issues a two-step instruction to Laura: first, to touch the smartphone's camera (l.42), then, to place her hand on the bed "right beneath where the camera would be" (l.43). Laura complies with the first part of this instruction by putting her hand below the phone's camera (l.44, Figure 5, left), and then on the bed beneath the camera (l.45, Figure 5, right). Morgan then produces a positive assessment of Laura's hand placement ("perfect", l.46). This assessment directly indexes what is observable by Morgan through the phone's camera. Morgan then requests Laura to move her hand to the right (l.47), which Laura does immediately after (l.48).

This practice—placing a hand on the camera, then back onto the object being inspected in front of the camera lens—allows both participants to calibrate their senses: the sighted perception of Morgan becomes aligned with Laura's tactile and haptic perception of her environment. Through this method, Laura's sense of 'where the camera is' becomes accessible to her sighted partner. Laura's understanding of the camera's position (and therefore of Morgan's viewpoint) is produced as a public, shared, resource and materialized in Laura's hand placement on the bed. In other words, in this part of the fragment and beyond, the distribution of the perceptual work [129] emerges through the routinized strategies initiated by the sighted co-participant.

*4.3.3 Discretizing the 'Same' Feature with Different Senses*

In the closing moments of the interaction (Fragment 1.4, Figure 6), Laura and Morgan agree that their visual and tactile perceptions concern the *same* spot. This final agreement—that they have, in fact, located the stain—is achieved by aligning Laura's tactile sensations with Morgan's visual indication of the stain's location.

```
47 MOR =okay. (.) now move your hand to the right
48     (0.4)*%(0.6)
   lau   -->*moves hand right-->
   lau       %shifts body and camera to the right-->
49 LAU okai/
50     (0.2)*(0.3)
   lau   -->*
51 MOR right%# there.
   lau   -->%
   fig       #fig.6 (left)
52     (1.1)
53 LAU yea*h?#
   lau    *scratches blanket with her fingers-->
   fig        #fig.6 (right)
54     (0.9)
55 MOR yup there's a little spot right there.
56     (0.2)
57 LAU o:kay! (0.6) h (0.3) perfect (0.3) *thank you so much!
```

Fragment 1.4. Laura and Morgan reach agreement on the position of the stain.

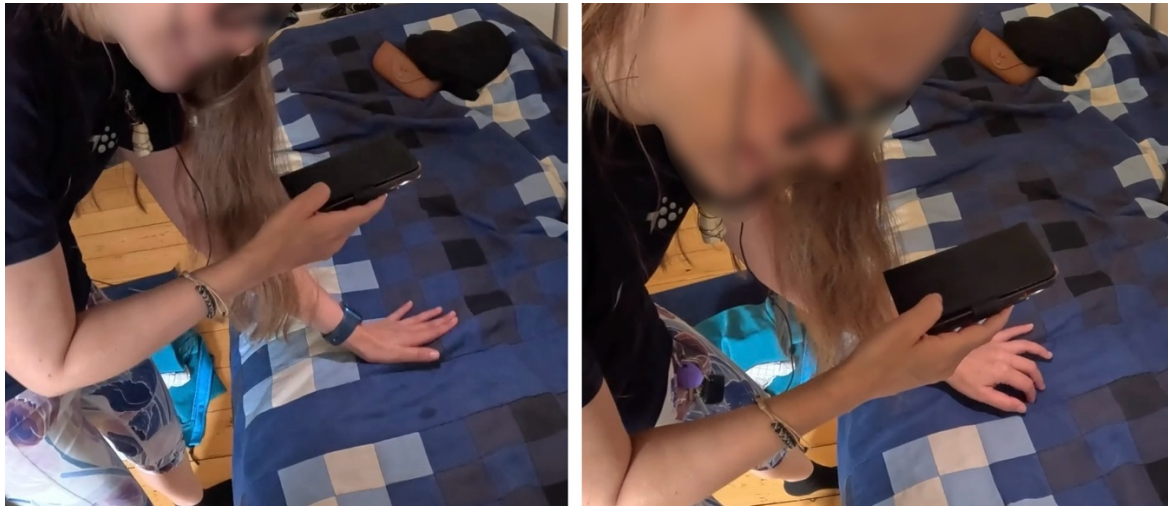

Figure 6. After moving her hand towards the right in response to Morgan's request (l.47), Laura holds the phone over her hand flat on the bed (left). After Morgan indicates the presence of a spot "right there" (l.51), Laura pinches the bed with her fingers (right).

This scene occurs after Morgan and Laura established a mutual point of reference (see Section 4.3.2 above)—visual and tactile. Morgan then instructs Laura to move her hand to the right (l.47). Laura complies by shifting her hand, body, and camera to the right (l.48), and by producing an agreement token ("okai", l.49). As Laura finishes repositioning herself, Morgan goes on to index a feature of their mutual environment ("right there", l.51, Figure 6, left). This "right there" orients both to what is currently visible to Morgan and to what is now tactilely available to Laura. Laura's interpretation of Morgan's turn is manifested in her subsequent scratching of the blanket at this location (l.53, Figure 6 right): in doing so, Laura connects (and displays connecting) Morgan's verbal account of the spot being "right there" to her own current tactile perception of the bed blanket. Laura's action is then ratified by Morgan as a successful identification of the correct spot ("yup there's a little spot right there", l.55). Morgan's indexical "right there" aligns Laura's tactile and Morgan's visual sensations as experiences of the same object [50]—the "little spot" (l.55). Through these practices of scratching and naming, Laura and Morgan accomplish a common "perceptual grounding" [95,175] of the stain, with its differentially perceived properties (its texture, color, shape, etc.). In a different theoretical vocabulary, these seemingly mundane practices resolve the ordinary yet critical issue of reconciling Laura and Morgan's perception of the world. They establish "sensory correspondences" [24] so as to form a world in common [114].

## 5 CO-INSPECTING A BLANKET WITH CHATGPT MULTIMODAL "VOICE MODE"

### 5.1 A One-Sided Distribution of the Work to Distribute Perception

In complete contrast to what Morgan accomplishes in the previous fragment, the contributions of ChatGPT multimodal "voice mode" (abbreviated as GPT in the following sections and transcripts) do not offer Laura resources to 'go on' with the inspection sequence. Additionally, they are systematically designed as what [162] term pre-closings: that is, turns that project the imminent closing of an interaction while providing participants with the option to confirm or extend the sequence. This concluding design of GPT's turns thus places the burden on Laura to keep the inspection sequence going and to provide GPT with new information about the bed.

```
31 LAU we+ll:%            (.)      +*what about uhm:: (0.2) % over here?
   lau       %orients torso and phone toward right of the bed%
   lau   +transfers phone right hand+
   lau                                  *scans bed with left hand-->
32     (0.2)
33 LAU +can you* see         + (.) if i do   *this like more closely.
   lau +lowers phone towards bed +holds phone still-->
   lau        *pushes clothes out of camera range*
34     (1.8)
35 GPT yes i can see that area a lot closer now and: (.) it %still looks clean! i
   lau                                                      %moves left-->
36     don't see any:* stains or marks so >i think you're good to* go.<
   lau            *-----------pushes clothes away-----------*
37     (0.4)
38 LAU what about here?
```

Fragment 2.1. Laura modifies her positioning and the camera angle before asking ChatGPT multimodal "voice mode" (GPT) for a new visual account

An example of the work accomplished by Laura is visible in Fragment 2.1 above. This fragment starts as Laura orients the camera towards the right part of the bed (l.31) then lowers it (l.33), thereby shifting from a macro-view, high above the bed, to a micro-view, closer to the blanket (l.33). She then asks GPT for an account of what it sees from this new position ("over here", l.31) and distance from the bed ("if I do this like more closely", l.33). GPT states that it can now see this part of the bed more closely (l.35), that this part still looks clean (l.35), and that it does not see any stain (l.36). This account is phrased as a conclusive contribution: it ends with an expression that projects the closing of the sequence ("you're good to go", l.36). This expression is uttered at a quicker pace than the rest of the sentence and ends with a falling intonation. Consequently, this turn from GPT offers no prosodic cues that would project a response or invite further action. However, while GPT provided this account, Laura was moving to her left (l.35) and pushing clothes away (l.36). In doing so, she was configuring a new perspective for GPT, about which she subsequently requests a novel account ("what about here", l.38). GPT is thus given new perceptual data to work on, and both participants move on with the inspection sequence. In sum, Laura systematically initiates the recurring modifications she makes to GPT's perspective. These modifications to GPT's field of view are neither instructed nor advised by GPT: the voice agent issues no requests to improve its view, nor does it support Laura as she incrementally adjusts the camera's perspective over the bed.

This "work to make technology work" [34,154] is achieved by Laura without any guidance from GPT. The main burden of the activity at hand (perceiving and organizing the environment to be perceived) falls almost entirely on Laura [26,68] and is seldom shared by the voice agent. In other words, 'perceiving' is not a continuously co-produced action in Laura and GPT's interaction. Consistent with accounts noting the difficulty [45], or the impossibility [85], for a disembodied AI to adjust and collaborate effectively with embodied humans in material settings, GPT provides no resources that Laura can use to improve their mutual perception of the object being inspected.

### 5.2 No Actions Contingent on the Observable Setting

The most distinguishing feature of Laura and GPT's interaction is its orientation towards a step-by-step (rather than overlapping and multilayered) structure: GPT and Laura's conduct is constituted of "well-delimited adjacent actions" [121] rather than "continuous adjustments" [121]. In this strictly step-by-step organization, GPT never initiates turns contingent (and designed as contingent) on ongoing and emergent *observable* features of the setting—those provided by the smartphone camera's video stream. As we detail in Section 4, this stands in sharp contrast to the *Be My Eyes* volunteer's conduct. GPT does not react to Laura's embodied actions (such as moving towards the left of the bed—Figure 7, l.36) and to the new point of view they provide on the inspected object. GPT does not—and, given its current implementation, cannot (see Section 6.2.1)—initiate a response occasioned by observable features of the setting, unless it is prompted to do so.

This property of GPT's conduct is oriented to by the blind participant. Laura's actions continually manage this absence of ongoing responsiveness to the visual setting by GPT: Laura systematically produces the same steps to restart the inspection sequence after each utterance produced by GPT. As a result, the same sequential structure emerges over the interaction between Laura and GPT, of which the following fragment is an instance. It occurs near the start of the interaction between Laura and GPT. By this point, Laura has explained to GPT that she is searching for a stain on her bed, is unsure of its location, and has already asked GPT once what it could see on the bed.

```
33 LAU +can you* see if i do this+ #like more closel*y.
   lau +lowers phone towards bed +holds phone still-->
   lau          *-pushes clothes out of camera range-*
   fig                              #fig.8
34     (1.8)
35 GPT yes i can see that area a lot closer now and: (.)      1. Account (GPT)
       it %still looks clean! I don't see any
36     *stains or marks so >i think you're good to* go.<
   lau     %moves left-->                                     2. Re-configuration (LAU)
   lau *-----------pushes clothes away------------*
37     (0.4)
38 LAU what about here?                                       3. New request (LAU)
39     (0.2)+(1.2)
   lau    -->+holds phone still-->                            4. Steady stance (LAU)
```

Figure 7: Second occurrence of the four-step structure repeated after each visual account provided by GPT. This structure occurs four times during GPT and Laura's inspection of the blanket.

This structure unfolds as follows (See Figure 7):

1. **GPT produces a visual account.** GPT briefly reports what it sees and does not see. It does so without providing further instructions as to how to improve its perception of the blanket, or how to go on with the inspection sequence (l.35 and l.36).
2. **Laura reconfigures the setting.** As GPT responds, Laura repositions herself, reorients the camera, and reorganizes her immediate environment. This includes removing clothes or rags from the bed in a manner that makes this environment less cluttered and more 'examinable' for stains for the computer vision algorithm (l.36).
3. **Laura requests a new account from GPT.** Towards the end of her modification of GPT's perspective and environment, Laura asks GPT what it perceives in this renewed setting (l.38). These requests are phrased in a manner that indexes a shift either in the observable environment or in GPT's viewpoint, e.g., "what about (over) here?" or "what do you see now?".
4. **Laura holds the camera still.** As Laura utters a new request to GPT as to what it perceives, she holds her arm with the camera perfectly still, pointing it toward the bed (l.39, Figure 8). Meanwhile, while remaining outside the camera's field of view, her other arm occasionally rearranges non-visible parts of the environment. Laura keeps the phone in this steady position at least during the first seconds of GPT's response.

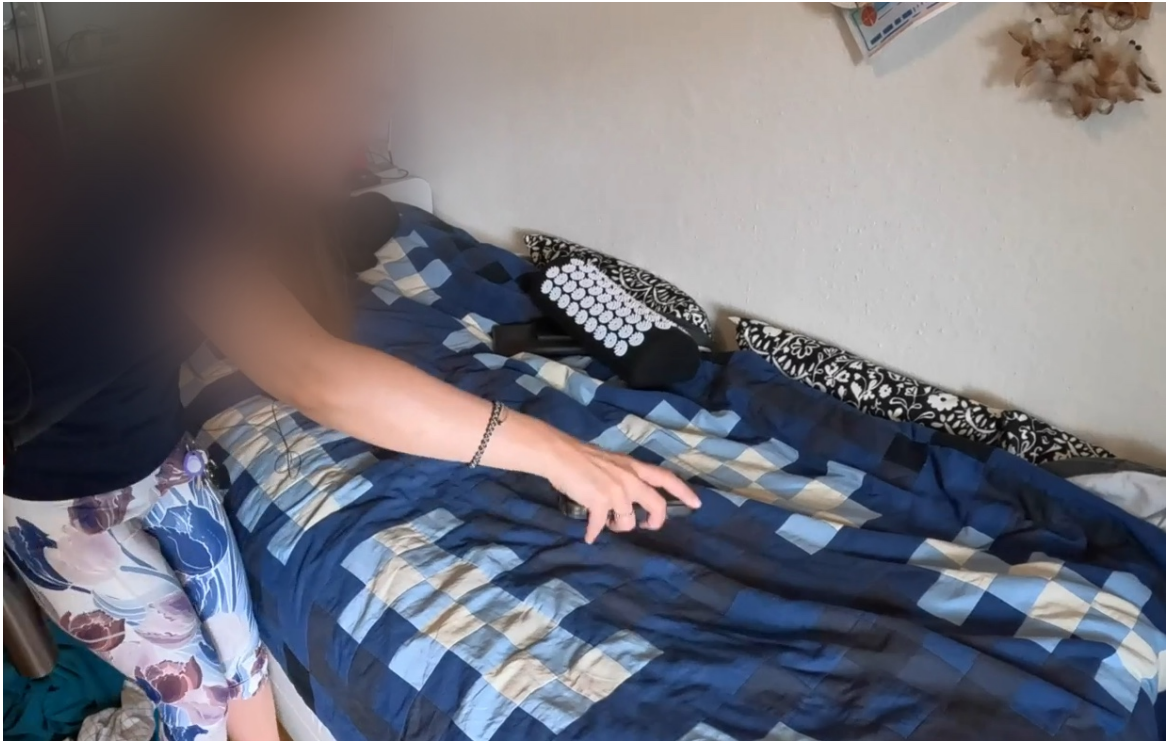

Figure 8. Laura stands still, holding her phone close to her bed, with the phone's camera oriented towards the blanket.

Overall, as often noted in HCI and HRI [33], Laura scaffolds [86,138] GPT by establishing conditions that enable it to make an adequate contribution. However, since GPT neither responds to the ongoing course of Laura's actions nor provides her with additional resources to 'go on', the inspection sequence must each time be started over by Laura, as she requests a new visual account from GPT. In another vocabulary, Laura and GPT do not achieve co-operative actions in Goodwin's [67] sense, where each participant's action responds and builds in real time on the other participant's hearable and/or embodied conduct.

## 6 DISCUSSION

### 6.1 Beyond 'Assistance': Similar Tasks, Widely Dissimilar Practices

Superficially, the two fragments analyzed above are of similar duration and contain a comparable number of turns. Yet, they differ markedly once the analysis shifts to the cooperative practices they feature. From this perspective, Laura and Morgan's activity was considerably more multilayered than the inspection sequence between Laura and the voice agent. Laura and Morgan produced a wide range of subtle adjustments and mutually responsive moves, extensively documented in prior EMCA research [121–123]. By contrast, Laura and GPT's interaction was organized as a strictly step-by-step process, akin to the rigid turn-taking observed in many voice-based AI assistants [2]. Furthermore, unlike what is recurrently observed in blind-sighted inspection sequences [50], this step-by-step progression in Laura and GPT's interaction relied alternatively on speech and sight, but never on tactile sensations. Added to this discrepancy—and possibly *because* of this absence of ongoing micro-adjustments—the attempt with GPT failed to locate the stain on the bed blanket. From another angle, Laura's options were partly constrained by GPT's conduct: it neither initiated nor provided Laura with resources to advance the interaction (such as instructions on how to improve its perception of the setting), nor did it build on Laura's own actions. It is in relation to the absence of these practices, in contrast to the human sighted assistant, that GPT's conduct can be labelled as less proactive. In particular, GPT did not:

1. Establish shared reference points across modalities (tactile, haptic, and visual) [130,144],
2. Provide the human with resources or explicit camera-work instructions to optimize its viewpoint [104]
3. Inform its actions by relying on the human's accounts of their current tactile or haptic sensations [24,131]
4. Adapt its utterances incrementally [3] (i.e., mid-turn) in response to the human's conduct or new information visible via the camera.

Overall, GPT's contributions were far less contingent on the local setting—and on Laura's conduct—than in the sighted volunteer's case. The interaction with GPT did not match the definition of "continuous user interaction scenarios, where user input is sustained and can receive AI feedback at any given moment for a more realistic and organic setup" [61]. As such, GPT's conduct offered fewer resources for Laura to distribute the work of sensing the environment together [120].

**6.2 How a Comparative Approach Can Inform Design**

*6.2.1 Situated Conduct and Its Technical Backdrop: What Voice Agents Do Not and Cannot Do*

This study has so far contributed to an empirical specification of remote sighted assistance and of one specific form of AI-powered assistance [62,102]. However, these insights gain in relevance when connected to the technological backdrop that shapes current voice agents' behavior. Indeed, Fragment 2 constitutes an exemplar of several documented flaws of LLM-powered voice agents. Notably, it displays the local realization of the well-established difficulty for an artificial agent to "share perception" [113] or to ground its actions in perceptual experiences [43]. It also illustrates the often-denounced tendency for LLMs to display overconfidence [28]—here, witnessable in GPT's categorical statements regarding the absence of a stain on a blanket. Finally, the limited range of turn-taking practices produced by GPT points to the limitations of the turn-taking models that underlie current voice agents' behavior [57,172,173,178]—independently of how good these voice agents' turn-design may otherwise be [183]. As noted by Addlesee and Papaioannou [3] in the case of voice assistants, voice agents do not currently modify their utterances incrementally during their very production. In other words, what GPT *does not do* in our fragment unsurprisingly connects with what LLM-based agents *cannot currently (reliably) do*.

In this situation, how can the strictly descriptive account this research has produced contribute to the design of language models that power current voice agents? Can pointing to the contrast in practices between human assistance and AI-powered assistance inform how large language models should be trained, and on which curated datasets? For example, should the training of these models prioritize datasets that feature certain cooperative methods—those identified by this research as crucial to the success of human collaboration within a specific task? HCI research has long emphasized the importance of situated action and contextual responsiveness in the design of interactive systems [42,177]. From this perspective, by pointing to the inadequacy of the local practices produced by LLM-powered voice agents, an EMCA approach hints at the incompleteness of the datasets and methods used to train these LLMs, as well as the limits of the turn-taking models that determine voice agents' turn-taking behavior. This practice-oriented line of critique is, in fact, already well established in research on conversational systems based on LLMs. For example, the training datasets of LLMs underlying artificial agents' (spoken) conversational practices—including their turn-taking behavior—have been called into question for being composed primarily of textual [127,178] and non-dialogical data [127,182].

*6.2.2 Current Multimodal Voice Agents Do Not Initiate Environmentally Occasioned Actions Based on Vision*

Above all, the preceding analysis reveals, on a moment-to-moment basis, how the absence of proactivity on the part of the artificial agent was consequential for the organization of the cooperative task—and had to be managed by the user. That is, a fundamental difference between the conduct of the voice agent and that of the human volunteer lay in the extent to which they accomplished practices typically classified as "proactive behavior" [198].

In particular, a pattern of conduct from GPT emerged as highly significant in the interaction with Laura: *unless Laura had just spoken*, the artificial agent never initiated a turn that publicly treated a feature of the setting (hearable, observable, or sequential) as relevant for the activity at hand (see Section 5.2). This conduct connects to a technical feature of current voice agents: they overwhelmingly rely on silence-based turn-taking models (as noted in [57,172,173,178]) and do not *initiate* noticings, comments, assessments, etc., about visible or audible features of the setting. Although ChatGPT multimodal "voice mode" uses computer vision, it cannot currently index visible features of the environment as publicly relevant by uttering a new turn in response to these features—rather than in response to an immediately prior human turn. As Umair, Sarathy, et al. [183] phrase it, "large language models know what to say but not when to speak". Agents relying on such turn-taking models will only produce such actions in second position, i.e., as responses to a human prompt. Even the more advanced continuous turn-taking models currently being developed [80,81,99,173,180] rely on spoken data in their probabilistic estimation of 'when' a turn could be relevantly initiated by a voice agent: they do not, yet, take into account observable features of the environment [29]. Although some early steps are being accomplished toward incorporating visual cues (e.g., [93,135]), those works focus on very specific embodied cues (e.g., gaze direction or head pose) rather than the broader observable setting.

Additionally, the voice agent did not (and, from a technical perspective, could not) *modify its turns during their very production* in response to newly observable features of the setting (e.g., Laura moving around, a stain appearing on camera, etc.). Ongoing reflexive adjustments—extensively documented by Goodwin [65,67] as a pervasive feature of human interaction [122]—were neither produced by nor accessible to the voice agent[4].

In sum, and in line with previous research [23,82], our analysis highlights that, for a multimodal voice agent to act as an expert 'guide' in assisting a blind person during an object-centered inspection sequence, it requires the capacity to continuously react to video data provided by the smartphone (or any other device)'s camera. To date, the proactivity framework as well as the concept of mixed-initiative interaction [74,133] are still meaningful for shedding light on the joint action sequences between Laura and ChatGPT multimodal "voice mode": these sequences were predominantly structured in response to the voice agent's limited capacity to take initiative.

**6.3 Looking Into the Future: Should Voice Agents Replicate All Human Cooperative Practices?**

*6.3.1 Environmentally Occasioned Actions Based on Vision as a Condition for Expert Collaboration*

In stark contrast with the interaction between Laura and ChatGPT "voice mode", vision-based environmentally occasioned actions were central to the collaboration between Laura and the *Be My Eyes* volunteer (see Section 4.2). Several of the volunteer's turns were recognizable *as responses to the observable setting*: their timing and phrasing indexed the appearance of a relevant detail (a potential location for the stain, Laura's hand, etc.) within the smartphone camera's frame. In this way, vision-based environmentally occasioned actions were critical to organizing an effective distribution of work between participants in investigating the location of the spot. These findings highlight that, if multimodal voice agents must act as human experts when cooperating in embodied tasks, they need to respond to more than human speech. Because even "turn taking is a mixed-initiative process" [23], an agent that cannot initiate new turns based on observable features of the environment is severely limited in the range of cooperation tasks it can undertake with humans [196].

---

[4] As canonically shown by Goodwin [66], utterances are commonly produced incrementally by their speakers—for example by adding "actually" to an ongoing sentence so that the addressee has time to shift their gaze towards the speaker [65].

In sum, in line with an important body of EMCA research, the analysis of our first fragment specifies how sighted-blind participants' cooperative practices recurrently rely on the ongoing and mutually responsive nature of their actions [24,50,92,170,171]. While acknowledging the classic argument that mediated 'collaboration' involving artificial agents should not merely imitate face-to-face human interaction—but may even exceed it [73]—we argue that, regardless of their design, voice agents will not surpass human sighted assistants so long as they lack access to a fundamental resource on which these human assistants rely: environmentally occasioned vision-based actions.

*6.3.2 Beyond the "Proactivity Dilemma": Environmentally Occasioned Actions and Their Moral Implications*

What if a voice agent could *initiate* some of the practices introduced by the *Be My Eyes* volunteer in Fragment 1? What if, during a collaboration with a human in cleaning a room, a voice agent produced requests like "move around so I can see better", without being prompted by the human? Or if it uttered "stop, let me see" as the human moved around—thereby reacting in real time to information provided by the smartphone's camera? Fundamentally, this question is part of the "proactivity dilemma" highlighted by Zargham et al. [198]. Any dialog system (among which voice agents) able to initiate new actions that, although potentially helpful, risk being intrusive or inappropriate [110,132,156,187]. Hence, beyond technical considerations, there appear to be ethical grounds for objecting to the accomplishment—by an AI—of practices displayed by the human sighted assistant in our first fragment.

Yet, the preceding analysis of two exemplar fragments underscores that the "proactivity dilemma" extends beyond the intrusiveness or inappropriateness of a dialog system's contributions. In the case of voice agents, this dilemma more fundamentally concerns who, in an ongoing interaction, should have the ability to steer the course of action in a particular direction without being explicitly prompted to do so. Indeed, in and through their environmentally occasioned actions (including those based on visible features of the environment), participants 'do' things [13,87]. In particular, they topicalize and draw attention to events or features within their immediate physical surroundings [69,190]. For example, the EMCA literature has recurringly documented how, through the practice of *noticing* [87], participants select and enforce which visible features of the environment (e.g., within the frame of the smartphone's video feed) are to be discretized and responded to [69,190]—by initiating a new turn.

Therefore, this debate about proactivity directly connects to broader practical and ethical questions about the latitude afforded to any form of AI in decision-making or collaborative processes [94,111,166]. Each turn produced by a voice agent during a mundane, ordinary assistive task—such as helping to find a stain—sets in motion longstanding questions about whether certain decisions should, or should not, be delegated to computers [191]. The production of environmentally occasioned actions based on vision raises not only computational but also interactional challenges, and it remains unclear whether voice agents should be endowed with the capacity to do so. Because initiating an action designed as "environmentally occasioned" *is* constituting certain aspects of the setting as features that deserve to be mentioned, monitored, or attended to, the broad set of practices encapsulated by this term unavoidably realizes a specific system of values and categories within an interaction—especially when what 'occasions' the contributions from a voice agent is not a human prompt. Enabling an artificial agent to initiate such actions (without merely replying to human requests) would represent a significant shift in the normative organization of talk [167] between humans and artificial agents. We argue that, through the production of environmentally occasioned actions by non-human agents, a fundamental change would occur in the typical distribution of roles in interaction—specifically, in who can legitimately display (by initiating a new turn or an embodied action) what features of the setting are to be treated as immediately relevant and as warranting a change in the current joint course of action.

*6.3.3 A Looming Ethical Challenge?*

The previous analyses highlight how even mundane, ordinary contributions such as noticings carry normative implications. These observations connect with a literature that frames formulations or descriptions as necessarily enacting a selection of aspects locally relevant to the situation at hand [39,157]. In an even broader perspective, this discussion intersects with works that analyze technological artifacts as *mediating* artifacts—that is, artifacts that shape one's interaction with the world [77,185] or, in a different theoretical framework, that transform the user's experienced world [141]. From this analytic standpoint, the ethical challenge is not to delineate *which human collaborative practices have no normative import*, but rather *which, if any, of these (inevitably) morally consequential forms of conduct should be designed*.

That is, if the technical challenges currently weighing on voice agents' turn-taking were to be solved (see Section 2.4.2), should these agents be designed to initiate or modify turns *solely* in response to specific features of their users' conduct—for example, exclusively in response to users' hand gestures or gaze direction? Unless explicitly prompted to do so, should these agents be configured not to react to any other feature of the environment, nor to initiate comments about the aesthetic, social, or practical value of users' surroundings? Yet should they nonetheless be required to initiate turns, or to modify their ongoing turns, in response to what they categorize as imminent threats—for example, a fast-approaching car suddenly appearing on a road? These questions are especially thorny in the case of technologies designed to support blind individuals. As noted in [4,8,112], blind people have fewer immediate resources to verify, and possibly resist, characterizations of their environment produced by artificial agents (such as voice agents relying on multimodal large language models). In this configuration, blind users may be more vulnerable to the ontologies [70] enacted by an artificial agent's descriptions of, or reactions to, their surroundings.

It is not the purpose of this article to determine the appropriate balance between collaborative efficiency and ethical caution. That is, the interactional perspective developed over the course of this work can only map out some of the empirical parameters that an ethical inquiry would need to consider. However, drawing on the previous insights, we suggest that ongoing debates in AI ethics should engage with this underexplored question. We point to the underestimated importance of this dilemma—namely, the extent to which an artificial agent can legitimately steer the course of an interaction by initiating new contributions—as it may manifest itself in everyday life, within ordinary and mundane interactions. We argue that, beneath their innocuous appearances, such ethically consequential situations are likely to become increasingly pervasive if voice agents' *continuous* turn-taking models—currently being developed [81,173]—advance to the point of becoming commercially viable.

# 7 LIMITATIONS: EXTERNAL VALIDITY OF THE FINDINGS

## 7.1 The (Statistical and Normative) Ordinariness of Situated Practices

The two fragments above are presented as heuristic illustrations of phenomena observed in two larger corpora (see Section 3.1.1), rather than as materials that would, in themselves, support general claims. Nevertheless, even when drawing on these broader data, the generalizability of our empirical findings is constrained by the small size of these corpora and is partially dependent on the specific configuration of the voice agent that was used. The previous analyses focused on offering situated qualitative insights rather than results intended for wider extrapolation. Hence, further work is needed to determine the extent to which the practices documented here are ordinary—both in a normative and a statistical sense—within assistive situations.

1. *Regarding the collaborative practices accomplished between Laura and the sighted volunteer* (Fragment 1), it is not possible to demonstrate their ordinariness on the basis of the present materials—that is, to confirm their status as typical and recurring components of situations of RSA. On the one hand, these practices are recurrent in our corpus of 20 BLV–sighted participant pairs engaged in object-centered sequences; for instance, uttering "right there" as the blind participant's gesture reaches the desired location, or various practices involved in establishing common reference points. As pointed out throughout the analysis, some of them are also extensively documented in the literature on interactions between blind individuals and sighted 'guides', as well as in the literature on instruction sequences [24,50,92,106,107,116,170,171]. On the other hand, any precise claim regarding the frequency of occurrence of these practices is made impossible by the limited size of our aforementioned corpus of BLV–sighted participant pairs, both in terms of participants and settings.
2. *Regarding the phenomena identified in human–voice agent interaction* (Fragment 2), a distinct dialogue system, used in another context, and trained on a differently constituted dataset, may have designed its turns otherwise. In such a case, it would have provided alternative constraints and opportunities for the blind participant, with whom other activities may have been co-constructed. For instance, the 'four-step' sequential structure detailed in Section 5.2 may not have occurred if the voice agent had offered Laura more resources to 'go on'—by articulating which camera position might enable it to obtain a better grasp of the environment, etc. In that configuration, the 'work to make technology work' accomplished by Laura may have taken a different form.

Further systematic comparisons between RSA and human–agent assistance could therefore be fruitful. Such comparisons should include a wider range of settings, devices (e.g., glasses equipped with cameras rather than a handheld smartphone), and assistive tasks, whether in controlled or naturalistic environments. In particular, additional research could examine whether the practices described above display and enact routines developed by expert participants over the course of similar situations, or whether they are also found among participants (blind or sighted) who have no prior experience with RSA. Finally, broader parameters identified in the literature on haptic exploration—such as the age of onset of blindness [100]—may also correlate with the specific methods we examined.

### 7.2 Insights Beyond the Local Implementation of the Voice Agent: The Conditions of Possibility for Specific Collaborative Practices

Despite the limitations just specified, a core general insight can be abstracted from the comparison between these two fragments. In effect, although we cannot demonstrate the ordinariness of the human collaborative practices investigated throughout this work, we can point out that, if a voice agent is to partake in these (in)ordinary practices, it must initiate or modify its turns in response to features of the observable environment. This insight is strictly analytic rather than empirical: the way in which our participants initiated and incrementally modified turns was a *condition of possibility* for their overarching collaborative practices (in creating points of reference, in distributing the perceptual work, etc.). Specifically, responding to visible details of the setting and tailoring one's contributions in real time are essential cogs in the machinery [158,168] through which the sighted human volunteer produced their conduct. Hence, regardless of the degree of refinement in a dialogue system's ability to *compose* its turns, this system's turn-taking model must at a minimum possess the properties described in Section 6.3.1 if it is to engage in the collaborative practices accomplished by the humans in our corpus. Following the same line of argument, these conditions of possibility also underpin similar practices, methods, and strategies documented in studies of interactions between sighted 'guides' and blind individuals [24,50,92,171].

## 8 CONCLUSION

This study contrasted how a blind participant searched for a stain on a blanket, depending on whether her assistant was a human remote sighted assistant or a multimodal voice agent. The comparison between two excerpts highlighted fundamental differences in how assistance was produced in each case, revealing that the success of human assistance relied on practices that the voice agent did not accomplish. In particular, the human remote assistant's conduct was characterized by:

1. The initiation of actions responsive to the observable environment;
2. The incremental modification of ongoing actions in response to the blind participant's evolving conduct;
3. A distribution of the work to improve each participant's perception of the environment.

These practices enabled human participants to mutually build on their specific resources and ultimately succeed in their task. On this basis, we argue that as long as multimodal voice agents cannot initiate actions in response to the video data available to them, they will miss a core resource routinely relied upon by human sighted assistants. The ability to initiate action in this way constituted the foundation of the collaborative practices we observed among our human participants. Specifically, the results of this empirical comparison underscore the continued relevance of the notion of proactivity for explaining the current limitations of multimodal voice agents in assistive tasks. Indeed, the effective practices on which the human assistance relied were overwhelmingly characteristic of the proactive behaviors that HCI and HRI researchers strive to develop in artificial agents.

Nevertheless, a collateral outcome of our study is to highlight, moment by moment, how initiating new actions (without responding to a human 'prompt') reorients the ongoing activity and imposes what is relevant to perceive or attend to. Thus, we suggest that it may not be ethically desirable for conversational agents to reproduce the full range of collaborative practices displayed by humans. Doing so would, in effect, place conversational agents in an unprecedented position to enact particular value systems at any point in an interaction.

# 10 APPENDIX

Transcription of talk follows Jefferson's transcription conventions [83]:
```
=         Latching of utterances
(.)       Short pause in speech (<200 ms)
(0.6)     Timed pause to tenths of a second
:         Lengthening of the previous sound
.         Stopping fall in tone
,         Continuing intonation
?         Rising intonation
°uh°      Softer sound than the surrounding talk
.h        Aspiration
h         Out breath
heh       Laughter
((text))  Described phenomena
```

Embodied actions were transcribed using Mondada's multimodal transcription conventions [117]:
```
**        Gestures and descriptions of embodied actions are
          delimited between:
++        two identical symbols (one symbol per participant)
ΔΔ        and are synchronized with corresponding stretches of
          talk.
*->       The action described continues across subsequent lines
-->*      until the same symbol is reached.
>>        The action described begins before excerpt's beginning.
-->>      The action described continues after the excerpt's end.
...       Action's preparation.
--        Action's apex is reached and maintained.
,,,       Action's retraction.
lau       Participant doing the embodied action is identified in
          small caps in the margin.
```

Abbreviations used in transcripts refer to the following dimensions:
```
LAU       Turn at talk from a participant (LAU, MOR, GPT)
```

```
lau       Multimodal action from a participant (lau, mor)
fig       Screenshot of a transcribed event
#         Position of a screenshot in the turn at talk
```